\documentclass[final,authoryear,3p,times]{elsarticle}

\usepackage{graphicx,color,colortbl}
\usepackage{enumerate}
\usepackage[colorlinks]{hyperref}
\usepackage{amsmath,amssymb,amsthm,bm}
\usepackage{multicol}
\usepackage[dvipsnames]{xcolor}
\usepackage{ulem}
\usepackage{tikz}
\usetikzlibrary{automata,positioning}
\usepackage{afterpage}
\usepackage{calc}
\usepackage{ifthen}
\usepackage{algorithm}
\usepackage{algpseudocode}
\usepackage{caption}
\usepackage{subcaption}
\usepackage{siunitx}
\usepackage{booktabs}

\edef\svtheparindent{\the\parindent}
\usepackage{parskip}
\parindent=\svtheparindent\relax

\newcommand{\CA}{\cellcolor{green!15}}
\newcommand{\CB}{\cellcolor{blue!15}}
\newcommand{\CC}{\cellcolor{orange!15}}

\newcommand{\boldface}[1]{\boldsymbol{#1}}  

\newcommand{\bfb}{\boldface{b}}

\newcommand{\bfk}{\boldface{k}}

\newcommand{\bft}{\boldface{t}}
\newcommand{\bfu}{\boldface{u}}
\newcommand{\bfv}{\boldface{v}}

\newcommand{\bfy}{\boldface{y}}

\newcommand{\bfA}{\boldface{A}}

\newcommand{\bfC}{\boldface{C}}

\newcommand{\bfF}{\boldface{F}}

\newcommand{\bfI}{\boldface{I}}

\newcommand{\bfK}{\boldface{K}}

\newcommand{\bfP}{\boldface{P}}
\newcommand{\bfQ}{\boldface{Q}}

\newcommand{\bfX}{\boldface{X}}

\newcommand{\bfbeta}{\boldsymbol{\beta}}

\newcommand{\bftheta}{\boldsymbol{\theta}}

\newcommand{\bfvarphi}{\boldsymbol{\varphi}}

\newcommand{\bfPhi}{\boldsymbol{\Phi}}

\newcommand{\calD}{\mathcal{D}}

\newcommand{\calK}{\mathcal{K}}

\newcommand{\calN}{\mathcal{N}}

\newcommand{\calU}{\mathcal{U}}

\newcommand{\calX}{\mathcal{X}}

\newcommand{\partderiv}[2]{\frac{\partial #1}{\partial #2}}

\newcommand{\Rset}{\mathbb{R}}

\newlength{\boxwidth}
\def\dd{\;\!\mathrm{d}}

\def\btheorem{\begin{theorem}}
\def\etheorem{\end{theorem}}
\def\blemma{\begin{lemma}}
\def\elemma{\end{lemma}}
\def\bproposition{\begin{proposition}}
\def\eproposition{\end{proposition}}
\def\bcorollary{\begin{corollary}}
\def\ecorollary{\end{corollary}}
\def\bdefinition{\begin{definition}}
\def\edefinition{\end{definition}}
\def\bexample{\begin{example}}
\def\eexample{\end{example}}
\def\bremark{\begin{remark}}
\def\eremark{\end{remark}}

\newcommand{\be}{\begin{equation}}
\newcommand{\ee}{\end{equation}}
\newcommand{\beq}{\begin{eqnarray}}
\newcommand{\eeq}{\end{eqnarray}}
\newcommand{\bem}{\begin{multline}}
\newcommand{\eem}{\end{multline}}
\newcommand{\ba}{\begin{align}}
\newcommand{\ea}{\end{align}}

\renewcommand{\figurename}{Figure~}
\renewcommand{\tablename}{Table~}

\journal{CMAME}

\begin{document}

\begin{frontmatter}

\title{Unsupervised discovery of interpretable hyperelastic constitutive laws}

\author[eth]{Moritz Flaschel\corref{cor1}}
\author[eth,tud]{Siddhant Kumar\corref{cor1}}
\author[eth]{Laura De Lorenzis\corref{cor2}}

\cortext[cor1]{These authors contributed equally.}
\cortext[cor2]{Correspondence: ldelorenzis@ethz.ch}

\address[eth]{Department of Mechanical and Process Engineering, ETH Z\"{u}rich, 8092 Z\"{u}rich, Switzerland}
\address[tud]{Department of Materials Science and Engineering, Delft University of Technology, 2628 CD Delft, The Netherlands}

\begin{abstract}
	We propose a new approach for data-driven automated discovery of isotropic hyperelastic constitutive laws. The approach is \textit{unsupervised}, i.e., it requires no stress data but only displacement and global force data, which are realistically available through mechanical testing and digital image correlation techniques; it delivers \textit{interpretable} models, i.e., models that are embodied by parsimonious mathematical expressions discovered through sparse regression of a large catalogue of candidate functions; it is \textit{one-shot}, i.e., discovery only needs one experiment - but can use more if available. The problem of unsupervised discovery is solved by enforcing equilibrium constraints in the bulk and at the loaded boundary of the domain. Sparsity of the solution is achieved by $\ell_p$ regularization combined with thresholding, which calls for a non-linear optimization scheme. The ensuing fully automated algorithm leverages physics-based constraints for the automatic determination of the penalty parameter in the regularization term. Using numerically generated data including artificial noise, we demonstrate the ability of the approach to accurately discover five hyperelastic models of different complexity. We also show that, if a ``true'' feature is missing in the function library, the proposed approach is able to surrogate it in such a way that the actual response is still accurately predicted. 
	
\end{abstract}


\end{frontmatter}


\section{Introduction}\label{sec:Introduction}
Data-driven approaches enabled by
machine learning tools and facilitated by the large availability of data through modern experimental techniques are raising a rapidly growing interest in computational solid mechanics. 
A main focus of recent investigations has been the possibility to \textit{bypass} or \textit{surrogate} the development of material models, based on the recognition that, unlike the other components of a mechanical boundary value problem (conservation laws and essential constraints), the constitutive modeling step is of empirical, non-epistemic nature. As follows, we attempt a brief overview of the main available approaches for reversible material behavior (elasticity).

\textit{Bypassing} material modeling altogether is the core idea of the (constitutive) model-free data-driven computing paradigm \citep{kirchdoerfer_data-driven_2016,ibanez_data-driven_2017}. In these approaches, a boundary value problem is solved by retaining conservation laws and essential constraints,
and substituting the constitutive model with the direct use of data. The pioneering
approach in \cite{kirchdoerfer_data-driven_2016} results in a data-driven solver which aims
at assigning to each material point the state satisfying the conservation
laws that is closest to the data set. The first demonstrations focused
on linear elasticity \citep{nguyen_data-driven_2018,conti_data-driven_2018,kirchdoerfer_data-driven_2018} but recent attempts extend it to geometrically
nonlinear elasticity \citep{nguyen_data-driven_2018}, general elasticity \citep{conti_data-driven_2018},
elastodynamics \citep{kirchdoerfer_data-driven_2018}, inelasticity \citep{eggersmann_model-free_2019}, 
and fracture mechanics \citep{carrara_data-driven_2020}. 
In \cite{ibanez_manifold_2018}, the phenomenological constitutive model is replaced with an 
experimental constitutive manifold, which is reconstructed from data using manifold
learning methods. In a subsequent investigation focusing on elasticity
\citep{ibanez_data-driven_2017}, the data are used to identify a polynomial approximation
of the strain energy density. Recently, this approach has been extended
to a setting that fulfils the thermodynamics principles by construction
\citep{gonzalez_thermodynamically_2019}. More recent contributions pursue a hybrid approach where
data are used to construct automatic corrections to existing hyperelastic
and elastoplastic models \citep{ibanez_hybrid_2019,gonzalez_learning_2019}. 

A second stream of investigations has focused on the idea of \textit{surrogating} constitutive models, e.g., through piecewise polynomial interpolation \citep{sussman_model_2009,crespo_wypiwyg_2017} or artificial neural networks (ANNs).
The first use of ANNs to encode material models dates
back to the 1990s \citep{ghaboussi_knowledgebased_1991}. The approach consists in replacing
a constitutive relation with the
training of an ANN based on data. The trained ANN encodes the relation
between input and output variables (in the simplest case, strains
and stresses, respectively) and can then be used as a black-box substitute of the
constitutive equations. The method was further developed for a variety
of applications \citep{ghaboussi_knowledgebased_1991,lefik_artificial_2003,shen_finite_2005,Mozaffar-plasticity-2019,kumar_metamaterials_2020}. 
Over alternative approaches to encode constitutive relations such as interpolations using piecewise linear functions, polynomial functions, or radial basis functions (RBF), ANN and especially deep ANN possess several advantages \citep{huang_learning_2020}: 
 using appropriate regularization, they have good generalization properties to unseen data (albeit they show poor extrapolation beyond the  interpolatory hull of the training data);
they work well even for complex, highly inhomogeneous or anisotropic distribution of training points, 
i.e., dense along certain directions and sparse along others. 

The above two categories of approaches are clearly \textit{uninterpretable} from the standpoint of the material model, 
as they suppress it or encode it with black-box ANNs, respectively. 
This limits the insight that they can bring towards the physical understanding of the material behavior based on experimental observations. 
A further crucial aspect is that the vast majority of the approaches of both categories are rooted in \textit{supervised
learning}, i.e., they need data consisting of input-output pairs and thus must rely on
stress data. Datasets may come from experimental measurements, or be generated
computationally by simulations at the lower scales within multiscale
approaches \citep{hashin_analysis_1983,yuan_toward_2008,wang_multiscale_2018}. For mechanical
tests, stress data are only obtainable in the simplest situations,
e.g. uniaxial tensile or bending tests. The comprehensive observation
of strain-stress relations relying on these tests is nearly impossible.
Multiscale simulations can generate training data sets, but their computational
cost is still unaffordable. Recognition of this issue is very recent in both the above categories of approaches.  
Within the data-driven computing paradigm, it motivated the development of the data-driven identification
method \citep{leygue_data-based_2018,dalemat_measuring_2019}, which formulates the inverse problem associated
to the approach in \cite{kirchdoerfer_data-driven_2016}. By using only displacement field and
load measurements, a set of admissible material strain-stress states
is recovered with no assumption on the constitutive equation. Within the stream of research on surrogating constitutive models with ANNs, 
recent attempts to use only displacement and
global force data have been performed in \citep{tartakovsky_learning_2018,huang_learning_2020,haghighat_deep_2020}. These approaches bear
relation to the so-called physics-informed neural networks (PINN) \citep{raissi_physics-informed_2019}.
The training of PINNs is performed with a cost function that, in addition
to data, includes the governing equations, initial and boundary conditions. 
However, for the time being, these methods are limited to only very simple cases (constitutive models of known form with unknown parameters or unknown constitutive models but for one-dimensional cases) and very simple 
ordinary and partial differential equations (ODEs and PDEs) -- e.g., common contemporary benchmarks include Burgers'equation, Darcy flow, and the advection-diffusion equation. In contrast, the ``constitutive ingredient'' (to be discovered) is much more complex and highly nonlinear in  most solid mechanics PDEs, even in hyperelasticity.

A recent inspiring stream of research initiated by the physics community focuses on automated discovery 
of physical laws and governing equations \citep{schmidt_distilling_2009}.
These approaches are interpretable as
they lead to closed-form expressions of the governing ODEs and PDEs for the investigated systems.
A breakthrough was achieved in \cite{brunton_discovering_2016} by departing from
a high-dimensional library of nonlinear candidate functions and using
sparse regression to only select the dominant items, thus uncovering parsimonious
governing ODEs. The approach has drawn tremendous attention \citep{loiseau_constrained_2016,zhang_robust_2018,champion_data-driven_2019,hoffmann_reactive_2019,lai_sparse_2019,wu_numerical_2019,huang_data-driven_2020,schaeffer_extracting_2018-1, wang_discovery_2020}. An important extension has addressed
the data-driven discovery of PDEs with time and spatial coordinates
as variables \citep{rudy_data-driven_2017,schaeffer_learning_2017}. Crucial improvements have
been obtained by substituting numerical differentiation with automatic differentiation
\citep{baydin_automatic_2017} and smoothing noisy data with neural networks \citep{berg_data-driven_2019,both_deepmod_2019,chen_deep_2020}. 
These approaches do not limit in any way the form of the sought ODE or PDE, but only depart from the 
assumption of a possibly very large catalogue of candidate component functions. 
Thus, apart from the choice of this catalogue, they do not leverage domain knowledge in the form of constraints 
stemming e.g. from conservation laws, thermodynamics, or symmetry requirements 
and seek the direct determination of the entire PDE. The consequences are two-fold: 
if directly applied to solid mechanics, these approaches may be more challenging than needed, 
and they may fail to comply with the epistemic requirements based on physics which amount to essential components of solid mechanics boundary value problems.

In this paper, inspired by the above ideas and tools, we propose a new approach which aims neither to bypass nor to surrogate, but rather to \textit{automatically discover} constitutive material models, focusing on isotropic hyperelasticity in this first investigation. The approach delivers \textit{interpretable} models, i.e., models that are embodied by parsimonious mathematical expressions based on sparse regression of a large catalogue of  candidate functions, and is based on \textit{unsupervised learning}, i.e., it takes as input only experimentally measurable data in the form of full-field displacements, as obtainable e.g., from digital image correlation (DIC) techniques, and global force data delivered by mechanical testing machines. Unsupervised discovery is achieved by enforcing the constraints stemming from physics, including balance of linear momentum and additional constraints for the elastic strain energy. Finally, using machine learning jargon, our approach is \textit{one-shot}, meaning that discovery only needs one experiment - but can obviously use more if available. This is in contrast to conventional material model calibration techniques as well as supervised learning methods, that require several different experiments (e.g., uniaxial or biaxial tension, shearing, torsion, randomized strain paths, etc.).

\section{Unsupervised discovery of hyperelastic constitutive laws}

\subsection{Problem setting} \label{sec:problem_setting}

Consider a reference domain $\Omega\in\Rset^2$, with boundary $\partial\Omega$, subjected to a quasi-static mechanical test.\footnote{While we consider a two-dimensional setting here motivated by DIC experiments, the subsequent derivations are also applicable to the three-dimensional case (which is not our focus here).}
Dirichlet and Neumann boundary conditions are applied by the testing machine on $\partial\Omega_u\subseteq \partial \Omega$ and $\partial\Omega_t = \partial\Omega\backslash\partial\Omega_u$, respectively. Without loss of generality, we assume the experiments delivering the data to be conducted in displacement control, so that  the Neumann boundary conditions are homogeneous and the forces applied by the testing machine to the specimen correspond to the reaction forces at the Dirichlet boundary. 

Let $\calX=\{\bfX^a\in \Omega : a=1,\dots,n_n\}$ denote a set of coordinates of $n_n$ nodes in the reference configuration for which the displacements $\calU=\{\bfu^a\in \Rset^2 : a=1,\dots,n_n\}$ are known. In the present two-dimensional setting, this discrete full-field displacement dataset is assumed to be known through DIC measurements. The extension to the three-dimensional setting would be conceptually straightforward and would require full-field displacement data e.g. from computed tomography combined with digital volume correlation.  Additionally, the net reaction forces on portions of $\partial\Omega_u$ (corresponding to, e.g.,  each loaded side of the specimen) are assumed to be known and emulate the measurements from load cells.

\begin{figure}[!t]
	\begin{center}
		\includegraphics[width=\textwidth]{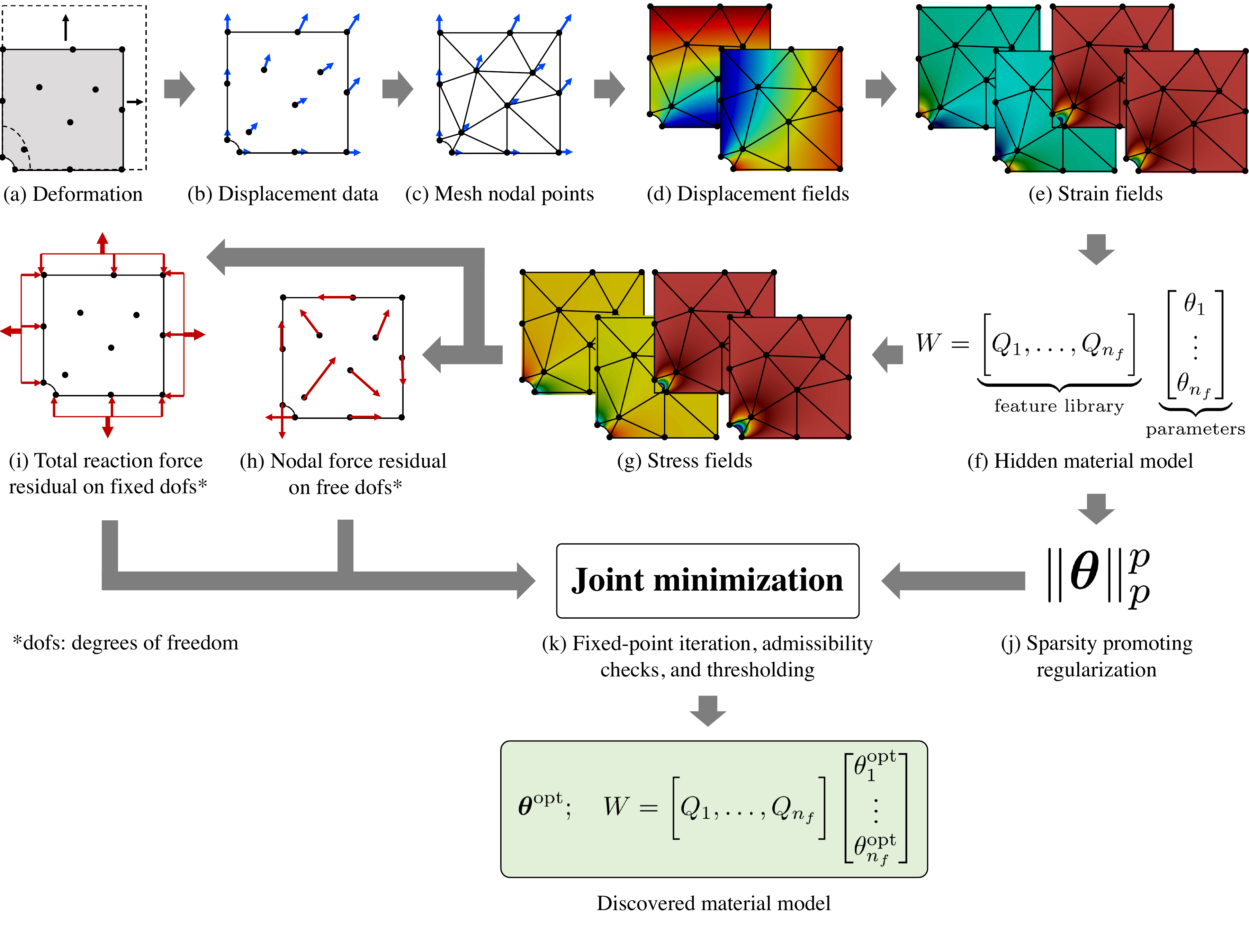}
		\caption{Schematics of the unsupervised algorithm for discovering interpretable hyperelastic constitutive laws. Starting from a deformed body under loading (a), the point-wise displacement measurements (b) are interpolated by constructing a  finite element mesh (c). The resulting continuous displacement field (d) is differentiated to obtain the strain field (e). The material  strain energy density~$W$ is formulated as a linear combination of a large catalogue of nonlinear features~$\bm{Q}$ and unknown material parameters~$\bm{\theta}$ (f). The derivative of the strain energy density with respect to the strain field yields the stress field (g). A joint optimization problem (k) is formulated to find the unknown material parameters $\bm{\theta}$ such that the weak form of the linear momentum balance in the bulk material (h) and the reaction force balance on the Dirichlet boundaries (i) are satisfied. The joint optimization problem (k)  also includes a  sparsity-promoting regularization (j) to yield a parsimonious and interpretable material model (characterized by $\bftheta^{\text{opt}}$).}
		\label{fig:schematics}
	\end{center}
\end{figure}

Armed with these data, our objective is to discover the underlying  constitutive law of the unknown material in one-shot (i.e., we only consider one  loading experiment). We note that the experiment must be complex enough to activate a diverse range of deformation modes. We  demonstrate in Section~\ref{sec:benchmarks} that biaxial tension of a plate with a hole suffices for this purpose. For the scope of this initial work, our key assumptions are that the material is homogeneous, isotropic, and hyperelastic. {\figurename\ref{fig:schematics} illustrates a step-by-step schematics of the proposed unsupervised algorithm, whereas a detailed description is presented in the following sections.}

\subsection{Displacement field approximation via finite element mesh}\label{sec:displacement_field}

 To facilitate derivatives and integrals of the field quantities, we associate the grid at which the displacements are known to a  finite element mesh with the nodal set $\calX$.  The displacement field is then approximated as
\begin{align}
\label{eq:discretization_u}
\bfu(\bfX) = \sum_{a=1}^{n_n}N^a(\bfX) \,\bfu^a,
\end{align}
where $N^a:\Omega \rightarrow \Rset$ is the shape function associated with the node of reference coordinate $\bfX^a$. The deformation gradient is given by
\begin{align}
\label{eq:discretization_F}
\bfF(\bfX) = \bfI+ \sum_{a=1}^{n_n}\bfu^a \otimes \nabla N^a(\bfX),
\end{align}
where $\nabla$ denotes the gradient operator with respect to the reference coordinates.

\subsection{Material model library}
\label{sec:material_model}

The objective is to find a constitutive law for the unknown material under consideration, e.g. a relation between the deformation gradient $\bfF$ and the first Piola-Kirchhoff stress tensor $\bfP$.  Since the material is hyperelastic, our objective reduces to discovering the underlying strain energy density $W(\bfF)$ from which the stress can be derived as $\bfP=\partial W/\partial \bfF$. Moreover, the objectivity requirement is a priori satisfied if we assume $W$ to be a function of the right Cauchy-Green deformation tensor  $\bfC=\bfF^T\bfF$. Finally, in the scope of isotropic materials, the strain energy {density} can be further reduced to a function of the invariants of $\bfC$ as
\begin{equation}\label{eq:strain_energy_density}
W(\bfC) \equiv W(I_1(\bfC),I_2(\bfC),I_3(\bfC)),  \qquad \text{with} \quad  
I_1(\bfC) = \text{tr}(\bm{C}), \quad 
I_2(\bfC) = \frac{1}{2}[\text{tr}(\bm{C})^2-\text{tr}(\bm{C}^2)], \quad
 I_3(\bfC) = \text{det}(\bm{C}).
\end{equation}

We consider a large feature library $\bfQ:\Rset^3\rightarrow \Rset^{n_f}$ of $n_f$ nonlinear functions of $I_1$, $I_2$, and $I_3$  whose linear combination yields the strain energy density as
\begin{equation}\label{eq:ansatz}
W(I_1,I_2,I_3) = \bfQ^T (I_1,I_2,I_3) \,\bftheta,
\end{equation}
where $\bftheta\in\Rset^{n_f}$ are the unknown material parameters to be estimated. In principle, any  arbitrary  strain energy density can be represented by \eqref{eq:ansatz} if the  feature library is appropriately chosen. In the context of our work, we choose the following library
\begin{equation}\label{eq:features}
\bfQ (I_1,I_2,I_3)= 
\underbrace{\left[ (\bar I_1-3)^i(\bar I_2-3)^{j-i} : j\in \{1,\dots,N\}, i\in\{0,\dots,j\}\right]^T}_{\text{Generalized Mooney-Rivlin features}}
\oplus 
\underbrace{\left[ (J-1)^{2k} : k \in\{ 1,\dots,M\}\right]^T}_{\text{Volumetric deformation features}}
\oplus 
\underbrace{\left[ \log \left(\bar I_2 / 3\right)\right]}_{\text{logarithmic feature}},
\end{equation}
where $\bar{I}_1=J^{-2/3}I_1$, $\bar{I}_2=J^{-4/3}I_2$, $J=\det(\bfF)=I_3^{1/2}$, and $\oplus$ denotes vector concatenation. The choice of $N$ and $M$ determines the total number of features $n_f$. Throughout this work, $N=M=7$ is chosen, resulting in $n_f=43$ features. The first set of features in \eqref{eq:features} is motivated from polynomial hyperelasticity laws such as those described by generalized Mooney-Rivlin models for rubber-like materials \citep{hartmann_numerical_2001,marckmann_comparison_2006,bower_applied_2009}. The second set of features allows capturing volumetric deformation in compressible elasticity. The third term is motivated from logarithmic features encountered in models such as those proposed by \citet{gent_forms_1958}. We  later exclude this term purposefully to test the generalization capability of our approach when the appropriate features are missing.

The components of the first Piola-Kirchhoff stress tensor are computed from \eqref{eq:ansatz} as
\be
\label{eq:stress}
P_{ij} = \partderiv{W(I_{1},I_{2},I_{3})}{ F_{ij}} = \partderiv{\bfQ^T(I_{1},I_{2},I_{3})}{F_{ij}}\bftheta.
\ee
The derivatives of the feature library $\bfQ$ with respect to the deformation gradient can be calculated by applying the chain rule. Owing to the severe nonlinearities in $\bfQ$, manual differentiation is  intractable. Therefore, we compute the derivatives via automatic differentiation (see \cite{baydin_automatic_2017} for a detailed review) that provides exact derivatives while bypassing the computational expense and accuracy issues of symbolic  and   numerical differentiation, respectively.

The above choice of the feature library is conducive to both physical and mathematical interpretability of the discovered models and, unlike in model-free and ANN encoding approaches, leverages decades of physical and phenomenological  knowledge in  modeling of hyperelastic materials, and in particular of rubber-like materials. For example, \eqref{eq:features} naturally admits  physics-based  models, e.g., the Neo-Hookean \citep{treloar1943} and  Isihara \citep{isihara} models which are derived via statistical homogenization of polymeric molecular chains. The model proposed by \cite{ArrudaBoyce1993}, based on Langevin statistics of polymeric chains, does not have a closed form expression (due to the presence of the inverse Langevin function) and, in practice, is often approximated by a Taylor series with polynomial features which are already present in \eqref{eq:features}. The feature library also admits phenomenological models such as those proposed by  \citet{biderman}, \citet{haines_wilson},  \citet{gent_forms_1958}, and many more. We refer the reader to the aforementioned references for physical interpretation of their respective features. The  library can be easily expanded to include new features if needed, possibly based on the prior knowledge of general and/or specific features of the material behavior, e.g. when dealing with biological tissues rather than rubber-like materials. Moreover, physical constraints --  e.g., material objectivity, material symmetries, stress-free reference configuration (i.e., $\bfP(\bfF=\bfI)=\bf0$),  can be satisfied automatically. In the present case, objectivity and material symmetry (isotropy) are ensured a priori by \eqref{eq:ansatz}, and a stress-free reference configuration is guaranteed by \eqref{eq:features}. Checking additional constraints \citep{ball_1976} such as boundedness, coercivity, and convexity  of the discovered strain energy density via (semi-)analytical methods is facilitated by the mathematically interpretable and parsimonious functional form. In general, leveraging such prior knowledge greatly reduces the solution space and makes the search for physically admissible models more tractable -- particularly in the unsupervised setting, where the absence of stress data typically renders the search highly ill-posed.

\subsection{Equilibrium constraints for unsupervised discovery}
\label{sec:weak_formulation}
On the basis of the interpolated displacement data (Section~\ref{sec:displacement_field}) and  the material model library (Section~\ref{sec:material_model}), the objective is now to find suitable material parameters $\bm{\theta}$, such that the physical constraints are satisfied, i.e., linear momentum balance\footnote{Note that the angular momentum balance is automatically fulfilled in the described setting.} is fulfilled both in the bulk material and at the boundary. This is a crucial step, as the enforcement of these physical constraints acts as a substitute for the availability of stress data and thus allows for unsupervised discovery. Assuming negligible body forces, the  weak formulation of linear momentum balance  in the reference domain $\Omega$ under quasi-static conditions is given by
\be\label{eq:weak_form}
\int_\Omega \bfP\colon\nabla \bfv \dd V - \int_{\partial\Omega_t}\hat \bft \cdot \bfv \dd S = 0, \quad \forall \quad  \text{admissible} \ \bfv,
\ee
where $\hat \bft$ is the surface traction acting on $\partial \Omega_t$ (zero in the present displacement-controlled setting, but non-zero for force-controlled experiments), and the test function $\bfv$ is admissible if it is sufficiently regular and vanishes on the Dirichlet boundary $\partial\Omega_u$. Note that we prefer the weak to the strong formulation of linear momentum balance because the double spatial derivatives required by the latter make it more sensitive to noise. This is also corroborated by the preferred use of the weak formulation in solving inverse problems based on full-field measurements, e.g., by applying the Virtual Field Method (see \cite{grediac_virtual_2008} for a review and \cite{tayeb_sensitivity_2020} for a recent application to hyperelasticity). Such methods show methodical parallels to the approach presented in this paper, however, they rely on the assumption that the form of the constitutive law is known \textit{a priori}.

Let $\calD=\{(a,i) : a=1,\dots,n_n;i=1,2\}$ denote the set of all nodal degrees of freedom. $\calD$ is further  split into two subsets of free and fixed (via Dirichlet constraints) degrees of freedom: $\calD^{\text{free}} \subseteq \calD$ and $\calD^{\text{fix}} = \calD \backslash \calD^{\text{free}}$, respectively. Using the same approximation basis as for $\bfu$ (Bubnov-Galerkin method), the test functions are approximated  as
\be
\bfv(\bfX) = \sum_{a=1}^{n_n}N^a(\bfX) \,\bfv^a, \quad \text{with}\quad v^a_i = 0 \ \  \text{if}\ \  (a,i)\in \calD^\text{fix},
\ee
The weak form \eqref{eq:weak_form} then reduces to 
\be\label{eq:weak_form_discretized}
\sum_{a=1}^{n_n} \bfv^a \cdot \left[ \int_\Omega \bfP \nabla N^a \dd V - \int_{\partial\Omega_t} \hat\bft  N^a \dd S\right] = 0.
\ee
Substituting the unknown material model \eqref{eq:stress} and noting that \eqref{eq:weak_form_discretized} must hold for all admissible $\{\bfv^a: a=1,\dots,n_n\}$, we obtain the following componentwise force balance equations at all free degrees of freedom:
\be
\int_\Omega \left(\partderiv{\bfQ^T}{F_{ij}}\bftheta\right) \nabla_j N^a \dd V - \int_{\partial\Omega_t} \hat t_i  N^a \dd S = 0,\quad \forall \quad (a,i)\in \calD^\text{free}.
\ee
where we use the Einstein summation convention. The integrals are computed via numerical quadrature on the mesh. {Under the assumption of material homogeneity, the material parameters $\bftheta$ do not depend on the spatial coordinates and can be written outside the integrals.} This results in a linear system of equations for $\bftheta$ which can be vectorized and assembled into the form
\be\label{eq:system_free}
\bfA^\text{free}\bftheta = \bfb^\text{free},
\ee
where $\bfA^\text{free}\in \Rset^{|\calD^{\text{free}}|\times n_f}$ and $\bfb^\text{free}\in  \Rset^{|\calD^{\text{free}}|}$. 

The material parameters $\bftheta$ must also satisfy the force balance on the Dirichlet boundary $\partial\Omega_u$. For the degrees of freedom under Dirichlet constraints, the reaction force must equal the internal force from the material, i.e.,
\be
\hat r^a_i = \int_\Omega  \left(\partderiv{\bfQ^T}{F_{ij}}\bftheta\right) \nabla_j N^a \dd V, \quad \forall \quad (a,i)\in \calD^\text{fix}.
\ee
In general, the reaction force is not known at every fixed degree of freedom. Instead, only the sums of the reaction forces  for subsets of the fixed degrees of freedom (each subset corresponding to one loaded side of the specimen and one coordinate direction) are known, emulating force measurements in an experiment. Let $n_{\alpha}$ be the number of these subsets in $\partial \Omega_u$. For $\alpha=1,\dots,n_\alpha$, let $\calD^{\text{fix},\alpha}\subseteq \calD^{\text{fix}}$ be the set of degrees of freedom under Dirichlet constraints {for which the sum of the reaction forces $\hat R^\alpha$ is known}, such that $\cup_{\alpha=1}^{n_\alpha} \calD^{\text{fix},\alpha} = \calD^{\text{fix}}$ and $\calD^{\text{fix},\alpha} \cap \calD^{\text{fix},\beta} = \emptyset$ for $\beta\neq\alpha$. The force balance for each subset is given by
\be
\hat R^\alpha = \sum_{(a,i)\in\calD^{\text{fix},\alpha}} \hat r^a_{i} = \sum_{(a,i)\in\calD^{\text{fix},\alpha}} \int_\Omega  \left(\partderiv{\bfQ^T}{F_{ij}}\bftheta\right) \nabla_j N^a \dd V, \quad \forall \quad  \alpha = 1,\dots,n_\alpha.
\ee
Note that each subset $\alpha$ only contains degrees of freedom corresponding to displacement components in one direction, so that the sum of the corresponding reaction force components is meaningful. Following the integration via numerical quadrature, this can be rearranged into additional $n_\alpha$ linear equations for $\bftheta$ of the form
\be\label{eq:system_fixed}
\bfA^{\text{fix}}\bftheta = \bfb^\text{fix},
\ee
where $\bfA^{\text{fix}}\in \Rset^{n_\alpha \times n_f}$ and $\bfb^{\text{fix}}\in \Rset^{n_\alpha}$.

The system of equations \eqref{eq:system_free} and \eqref{eq:system_fixed} are overdetermined and can be solved collectively in the least squares sense as
\be\label{eq:eqb}
\bftheta^{\text{opt}} = \arg \min_{\bftheta}
\bigg(
\underbrace{
	\left\| \bfA^\text{free}\bftheta - \bfb^\text{free}\right\|^2
}_{\text{linear momentum balance}}
+\lambda_r \underbrace{ 
	 \left\| \bfA^\text{fix}\bftheta - \bfb^\text{fix}\right\|^2
}_{\text{reaction force balance}}
\bigg),
\ee
with $\lambda_r>0$ as a hyperparameter weighting the contribution of the reaction force balance term. This weighting coefficient is important to ensure that both boundary and interior force balance terms are contributing to a similar extent to the value of the objective function. In the chosen spatial discretization/mesh (discussed in Section \ref{sec:benchmarks}), the number of boundary points is approximately two orders of magnitude smaller than the number of interior points. To ensure the similar influence of both sets of points in the optimization problem, $\lambda_r=100$ was chosen to compensate for the fewer equations arising from the boundary points and kept constant throughout the numerical experiments. In our experience, small variations in $\lambda_r$ do not have a significant influence on the solution to the minimization problem stated above. However, a violation of the balance equations at the boundary or in the interior is observed when choosing the order of magnitude of $\lambda_r$ too small or too high, respectively. The above minimization problem is also equivalent to solving
\be\label{eq:direct_solution}
\begin{aligned}
\bftheta^{\text{opt}} ={\bfA^{\text{eqb}}}^{-1}\bfb^{\text{eqb}} , \quad
\text{with}\quad 
&\bfA^{\text{eqb}} = {\bfA^{\text{free}}}^T\bfA^{\text{free}} + \lambda_r {\bfA^{\text{fix}}}^T\bfA^{\text{fix}},\\
&\bfb^{\text{eqb}} = {\bfA^{\text{free}}}^T\bfb^{\text{free}} + \lambda_r {\bfA^{\text{fix}}}^T\bfb^{\text{fix}}.
\end{aligned}
\ee

During an experiment, multiple full-field displacement datasets are usually available at different load steps and offer a richer information than from a single load step. Such information can be directly leveraged by additively combining the respective least square losses, i.e.,
\be
\bftheta^{\text{opt}} = \arg\min_{\bftheta} 
\sum_{l=1}^L
\left(
 \left\|\bfA^{\text{free},l} \bftheta - \bfb^{\text{free},l}\right\|^2
+\lambda_r  \left\|\bfA^{\text{fix},l} \bftheta - \bfb^{\text{fix},l}\right\|^2
\right),
\ee
where the superscript $l$ denotes the evaluation for  load step $l=1,\dots,L$. For the sake of notational clarity, the subsequent derivations are only shown for a single load step but can be straightforwardly extended to the general case.

\subsection{Parsimony and sparsity promotion}
\label{sec:sparsity}

Despite the optimization in \eqref{eq:eqb} being convex and apparently overdetermined in $\bftheta$, the nature of the problem is significantly ill-posed due to the unsupervised nature of the learning problem - i.e., the lack of stress field measurements. Both the feature library $\bfQ$ in \eqref{eq:features} and the coefficients $\bfA^\text{eqb}$  are highly nonlinear functions of the nodal displacements.  Consequently, the condition number of $\bfA^\text{eqb}$ is large and the optimization problem \eqref{eq:eqb} is highly sensitive to noise in displacement data measurements. Our numerical experiments (presented in Section~\ref{sec:results}) revealed that the optimization in \eqref{eq:eqb} is not sufficient to identify the correct underlying material model. To address this challenge, we  turn to additional sources of conceptual knowledge that can be leveraged to compensate for the lack of supervised data. 

The optimization problem in \eqref{eq:eqb} yields a dense solution, i.e., most entries in $\bftheta$ are likely  non-zero and many terms in the feature library \eqref{eq:features} will be included in the material model.  Previous works using supervised methods \citep{hartmann_numerical_2001} have shown that material models with a large number of material parameters are less interpretable, poor at extrapolation to unseen strains, and more likely to be physically inadmissible. In this light, we seek material models that are parsimonious with as few features as possible to explain the displacement data. 

To this end, we promote sparsity in $\bftheta$ by introducing an $\ell_{p}$  regularization in \eqref{eq:eqb} as
\be\label{eq:objective}
\bftheta^{\text{opt}} = 
\arg\min_{\bftheta} 
\left(
\left\|\bfA^{\text{free}} \bftheta - \bfb^{\text{free}}\right\|^2 
+\lambda_r\left\|\bfA^{\text{fix}} \bftheta - \bfb^{\text{fix}}\right\|^2 
+ \lambda_p\|\bftheta \|_p^p
\right),
\qquad \text{where} \qquad 
\|\bftheta\|_p = \left(\sum_{i=1}^{n_f}|\theta_i|^p\right)^{1/p}.
\ee
The hyperparameters $\lambda_p>0$ and $p\in[0,1]$  control the  degree of sparsity desired in $\bftheta$ and thus the selection of features that must be activated in the feature library \eqref{eq:features}. The $\ell_0$ regularization ($p=0$) is equivalent to penalizing the number of activated features and thus promotes parsimony and interpretability. However, the optimization becomes a combinatorial subset selection problem which is computationally intractable for a large number of features. Traditionally, the $\ell_0$ regularization is relaxed to the convex $\ell_1$ regularization ($p=1$) which promotes sparsity by zeroing out some coefficients  while shrinking the rest. Also known as LASSO (least absolute shrinkage and selection operator)  \citep{tibshirani_regression_1996}, $\ell_1$ regularization offers a good compromise between computational complexity and model interpretability. Within $[0,1]$, a smaller value of $p$ more aggressively promotes sparsity and interpretability at the cost of computational complexity (due to the higher degree of non-convexity in the penalty term) \citep{frank_statistical_1993}. In our numerical experiments, we observed $p=1/4$ to be a reasonable choice for the scope of our work.

\subsection{Numerical optimization strategy}

\subsubsection{Fixed-point iterative optimization}
\label{sec:fixed_point}

Due to the $\ell_p$ regularization with $0<p<1$, the objective function in \eqref{eq:objective} is non-convex and not continuously differentiable with respect to $\bm{\theta}$, thus, conventional gradient-based techniques such as gradient-descent and Newton (or Newton-like) methods do not perform well.  The development of new  algorithms for optimization problems including regularization terms of this type is an active  research area - see, e.g.,  \cite{wang_fast_2019} and references therein for a brief review. \citet{tibshirani_regression_1996} proposed a fixed-point iterative strategy for the $\ell_1$-regularized problem by solving a weighted $\ell_2$-regularized problem in each iteration. As follows, we generalize this idea to the $\ell_p$-regularized problem. 

The first optimality condition of the optimization problem \eqref{eq:objective} is given by
\be\label{eq:optimality}
2\bfA^{\text{eqb}}\bftheta + p\lambda_p  \left[
\text{sgn}(\theta_1)|\theta_1|^{p-1},\dots, \text{sgn}(\theta_{n_f})|\theta_{n_f}|^{p-1}\right]^T  = 2\bfb^{\text{eqb}},
\ee
where $\text{sgn}$ denotes the sign or signum function. This is a nonlinear system of equations and calls for  an iterative solution strategy. Let $\bftheta^{(k)}$ be the guess solution at the $k^\text{th}$  iteration. Within the $(k+1)^\text{th}$ iteration , \eqref{eq:optimality} is modified as
\be
2\bfA^{\text{eqb}}\bftheta^{(k+1)} + p\lambda_p  \text{diag}\left(
\left|\theta_1^{(k)}\right|^{p-2},\dots, \left|\theta_{n_f}^{(k)}\right|^{p-2}
\right)\bftheta^{(k+1)}  = 2\bfb^{\text{eqb}},
\ee
where we made use of $\text{sgn}\left(\theta_i^{(k+1)}\right)\left|\theta_i^{(k+1)}\right|=\theta_i^{(k+1)}$. Therefore, \eqref{eq:optimality} is reduced to a linear system of equations where the nonlinear terms are replaced by linear terms with weights based on the guess solution from the previous iteration, $\bftheta^{(k)}$. This further yields the fixed-point iterative scheme given by
\be
\bftheta^{(k+1)} \leftarrow \left[
\bfA^\text{eqb}
+\frac{p\lambda_p}{2} \text{diag}\left(
\left|\theta_1^{(k)}\right|^{p-2},\dots, \left|\theta_{n_f}^{(k)}\right|^{p-2}
\right)
\right]^{-1} \bfb^{\text{eqb}}.
\ee
To avoid division by zero,  whenever a coefficient $\left|\theta^{(k+1)}_i\right| < \epsilon_\text{tol}$ for some tolerance  $\epsilon_\text{tol}\ll 1$, the corresponding $i^\text{th}$ feature is removed from the  library $\bfQ$ and the fixed-point iteration scheme resumes with the remaining features only.
Further iterations are carried out  until the sequence $\{\bftheta^{(0)},\bftheta^{(1)},\bftheta^{(2)},\dots,\bftheta^\text{(k)},\bftheta^\text{(k+1)},\dots \}$ converges to a fixed point such that $\left\|\bftheta^{(k+1)} - \bftheta^{(k)}\right\|_\infty < \epsilon_\text{conv}$ for a convergence tolerance $\epsilon_\text{conv}\ll 1$.

The objective function landscape becomes highly non-convex with the introduction of the $\ell_p$ regularization. Depending on the choice of the initial guess $\bftheta^{(0)}$, the outlined fixed-point iteration scheme may converge to different local minima. For this reason, we perform multiple fixed-point iteration schemes in parallel, each with different randomly generated initial guesses. If the convergence criterion is not satisfied in a pre-specified maximum number of fixed-point iterations, the corresponding solution is discarded. From all the converged solutions, we finally choose the one which leads to the smallest value of the objective function \eqref{eq:objective}.

\subsubsection{Physical admissibility and the penalty parameter}
\label{sec:check_energy_requirements}

As discussed in Section \ref{sec:sparsity}, the hyperparameters $\lambda_p$ and $p$ determine the degree of sparsity of the solution. Having selected $p=1/4$ for our case (see Section \ref{sec:sparsity}), we now focus on the appropriate choice of $\lambda_p$ (penalty parameter). If $\lambda_p$ is too high, the resulting model will be too sparse and may not be rich enough to explain the displacement data. On the other hand, if $\lambda_p$ is too low, the optimization problem \eqref{eq:objective} becomes highly ill-posed and may yield physically inadmissible models.

\begin{figure}
    \centering
	\includegraphics[width=0.45\textwidth]{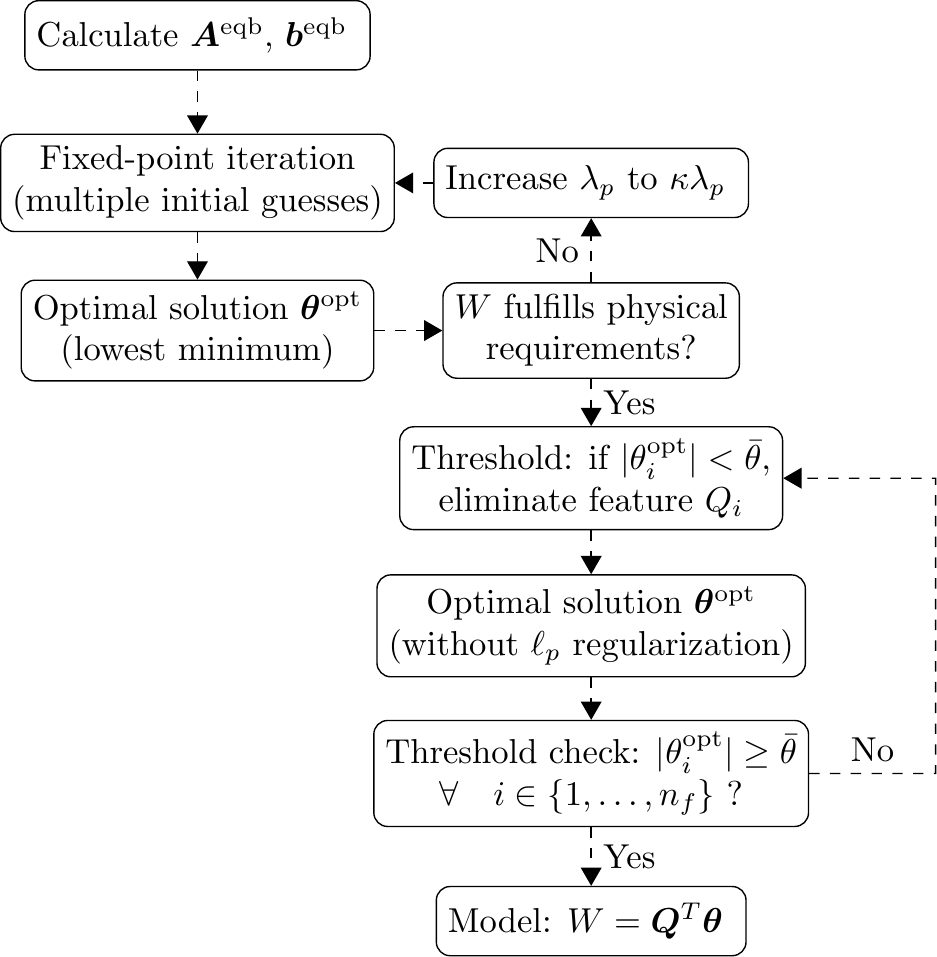}
	\caption{Schematics of the numerical optimization strategy. Note that the value of $n_f$ decreases during the iterative strategy as progressively more features are eliminated in the thresholding step.}
	\label{fig:sparsity_promotion_flowchart}
\end{figure}

In order to correctly estimate $\lambda_p$, we adopt an iterative strategy with the fixed-point algorithm nested inside it. We start with a  small initial value  of $\lambda_p = \lambda_p^0>0$. If the resulting model (obtained via the fixed-point algorithm described in Section \ref{sec:fixed_point}) is physically admissible (in the sense explained next), the model is accepted. Otherwise, $\lambda_p$ is multiplicatively increased by a factor $\kappa>1$, i.e., $\lambda_p = \kappa \lambda_p^0$. The process is repeated until a physically admissible model is obtained. By gradually increasing $\lambda_p$, we ensure that the resulting model is not too sparse or simple. \figurename~\ref{fig:sparsity_promotion_flowchart} describes the schematics of the optimization procedure.

To verify the physical admissibility of a model, we perform a series of empirical checks. An hyperelastic strain energy density must satisfy  $W(\bfF)\geq W(\bfI)=0$ for all physically admissible $\bfF$. While it is computationally intractable to verify this condition for all $\bfF$, we can easily perform the check for the deformation gradients observed in the domain $\Omega$. If the strain energy density is negative at any quadrature point in the finite element mesh, the material model is deemed inadmissible. 

Additionally, consider the following deformation paths
\begin{align}
&\bm{F}^{\text{UT}}(\gamma) = 
\begin{bmatrix}
1+\gamma & 0\\
0 & 1\\
\end{bmatrix},~
\bm{F}^{\text{UC}}(\gamma) = 
\begin{bmatrix}
\frac{1}{1+\gamma} & 0\\
0 & 1\\
\end{bmatrix},~
\bm{F}^{\text{SS}}(\gamma) = 
\begin{bmatrix}
1 & \gamma\\
0 & 1\\
\end{bmatrix},
\notag
\\
\label{eq:deformation_paths}
&\bm{F}^{\text{BT}}(\gamma) = 
\begin{bmatrix}
1+\gamma & 0\\
0 & 1+\gamma\\
\end{bmatrix},~
\bm{F}^{\text{BC}}(\gamma) = 
\begin{bmatrix}
\frac{1}{1+\gamma} & 0\\
0 & \frac{1}{1+\gamma}\\
\end{bmatrix},~
\bm{F}^{\text{PS}}(\gamma) = 
\begin{bmatrix}
1+\gamma & 0\\
0 & \frac{1}{1+\gamma}\\
\end{bmatrix},
\end{align}
which describe uniaxial tension (UT), biaxial tension (BT), uniaxial compression (UC), biaxial compression (BC), simple shear (SS) and pure shear (PS). Along these deformation paths, the strain energy density must increase monotonically, i.e., the following condition must hold true for all paths
\begin{align}
 0< W(\bm{F}(\gamma_i)) < W(\bm{F}(\gamma_j)) \quad \forall \quad \gamma_i<\gamma_j, \quad  \gamma_i,\gamma_j\in (0,\infty).
\end{align}
For computational tractability, the above inequality is tested for  a large number ($n_\gamma$) of samples of $\gamma_i$ and $\gamma_j$ from $(0,\gamma_\text{max})$ for some  $\gamma_\text{max}\gg 1$. If the inequality is violated along any of the deformation paths in \eqref{eq:deformation_paths}, the material model is rejected as unphysical. 

While these checks are only empirical in nature, in our experience they proved sufficient to constrain the solution space and obtain satisfactory results, see Section \ref{sec:results}. They also demonstrate that domain knowledge can be a very useful complement to a ``blind'' algorithm and help in determining the parameters of an otherwise difficult estimation. It is envisioned that a similar approach may involve the check of more rigorous mathematical properties of the strain energy density (e.g. polyconvexity or coercivity), which was not pursued in this work. Regardless of these checks, the mathematical interpretability and parsimony of the discovered models enable \textit{a posteriori} verification of e.g., polyconvexity with (semi-)analytical techniques, which is not feasible with black-box approaches including neural networks and model-free methods.

\subsubsection{Thresholding}

As a result of the $\ell_p$ regularization, the optimal solution obtained from the fixed-point iterations and physical admissibility tests  contains a large number of material parameters that are close or equal to zero. These ``inactive features'' can thus be permanently discarded from the initial library. This is performed through a simple thresholding operation analogous to the one used for supervised sparse regression by \cite{brunton_discovering_2016}, i.e., all parameters whose absolute value is below a threshold $\bar{\theta}$ are set to zero.  Subsequently, we can deactivate the $\ell_p$ regularization and solve the unregularized optimization problem \eqref{eq:eqb} for the remaining parameters. Note that the unregularized problem does not require an iterative technique and is solved directly using \eqref{eq:direct_solution} . If any of the resulting non-zero parameters is again below the defined threshold $\bar\theta$, the thresholding procedure is repeated until convergence is reached (see \figurename~\ref{fig:sparsity_promotion_flowchart} for schematics).

\section{Numerical benchmarks}\label{sec:benchmarks}

\subsection{Data generation} \label{sec:data_generation}
\begin{figure}[!ht]
	\begin{center}
		\includegraphics[width=0.5\textwidth]{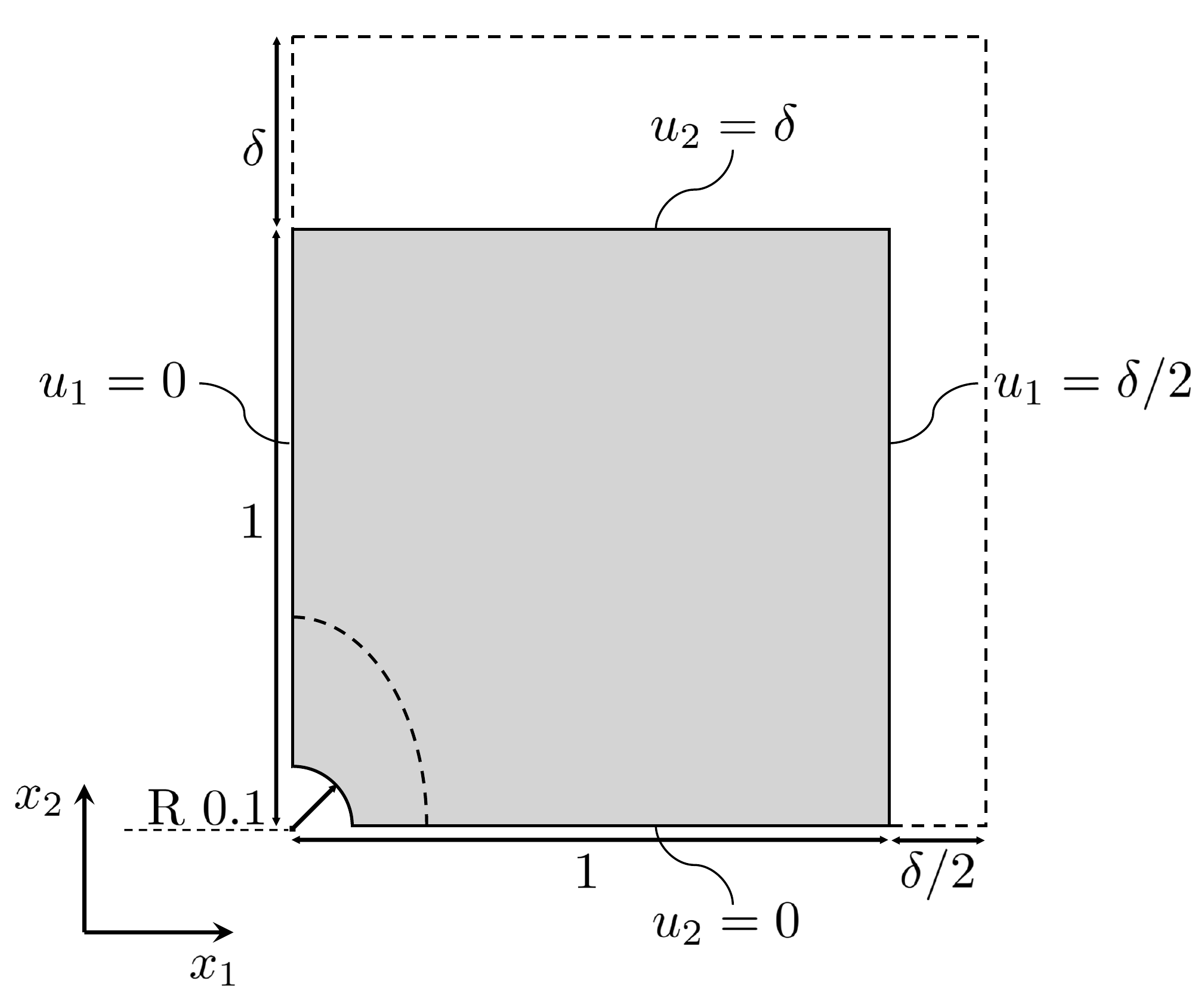}
		\caption{Geometry and boundary conditions of a quadrant of a plate with a hole under displacement-controlled asymmetric biaxial tension.}
		\label{fig:bcs}
	\end{center}
\end{figure}

For the scope of this work, we emulate DIC data with artificial data generated via the finite element method (FEM). We consider a hyperelastic square plate with a hole (as shown in \figurename\ref{fig:bcs}) under  displacement-controlled asymmetric biaxial tension with the loading parameter $\delta$. Due to the two-fold symmetry, only a quadrant of the plate is considered with symmetry boundary conditions on the left and bottom boundaries. All lengths and displacements are normalized with respect to the side length of the quadrant. The plate is assumed to be in plane strain conditions.\footnote{Note that, in plane strain, the right Cauchy-Green strain invariants are linearly dependent as $I_2=(I_1+I_3-1)$, thus the feature library can be written as a function of only two invariants, i.e., $\bfQ(I_1,I_3)$.} The domain is meshed with linear triangular elements. For the  prescribed boundary conditions (controlled via $\delta$), the nodal displacements are recorded from the FEM solution at a total of $L$ loadsteps. The total horizontal reaction force on the left and right boundaries, and the total vertical reaction force on the top and bottom boundaries are also recorded. We specifically choose this combination of geometry and loading, in-contrast to traditional bi-axial tension or torsion tests, because it leads to a strain field that is rich enough to solve the ill-posed problem of identifying the material model with no stress data and just one experiment.

The FEM simulations are performed with the following material models:
\begin{itemize}
\item {NH2}: Neo-Hookean solid with quadratic volumetric strain energy
\item {NH4}: Neo-Hookean solid with biquadratic volumetric strain energy
\item {IH}: Isihara solid \citep{isihara} with quadratic deviatoric strain energy
\item {HW}: Haines-Wilson solid \citep{haines_wilson} with cubic deviatoric strain energy
\item {GT}: Gent-Thomas solid \citep{gent_forms_1958} with logarithmic deviatoric strain energy
\end{itemize}
These models serve as benchmarks, as they are to be recovered through the exclusive use of nodal displacements and reaction force data using our approach. The exact strain energy densities are listed in Table~\ref{tab:material_models}.

In contrast to the synthetically generated data, real DIC data are unavoidably affected by noise in the measured displacement field. To emulate real data, we add artificial noise to the synthetic displacement data coming from the FEM simulations. As the noise in DIC measurements depends on the pixel accuracy of the imaging device, it would be unrealistic to assume the noise level to be proportional to the displacement magnitude. Instead, it is more realistic to assume the same absolute noise level (also referred to as noise floor) at every degree of freedom and for every load step, independently of the corresponding magnitude of displacement. We therefore add artificial noise to the displacement data such that
\be
u^{a,l}_i = u^{\text{fem},a,l}_i + u^{\text{noise},a,l}_i, \quad u^{\text{noise},a,l}_i \sim \calN(0,\sigma) \quad \forall \quad (a,i)\in \calD,~l\in\{1,\dots,L\},
\ee
where $u^{\text{fem},a,l}_i$ is the $i^{th}$ component of the displacement at node $a$ at load step $l$ obtained from the FEM, and $u^{\text{noise},a,l}_i$ denotes the noise added to it -- taken as a Gaussian noise with zero mean and standard deviation $\sigma>0$. A realistic estimate for $\sigma$  depends on the specimen size and the pixel accuracy of the measurement device. The noise floor typically ranges within $0.1-1\%$ of the pixel size (see, e.g., \cite{schreier_image_2009}) resulting in a reported strain measurement error of less than $10^{-4}$ relative to the domain length scale     \citep{bonnet_inverse_2005}. We hence consider two noise levels $\sigma \in \{10^{-4},10^{-3}\}$ (normalized with the plate quadrant length), which are (assuming sufficiently large specimen size and deformation) representative of respectively low and high noise levels encountered in modern DIC experiments.

Displacement measurements from DIC are often denoised by assuming spatial smoothness in the data. In previous works on data driven discovery of governing laws, denoising has been performed either as a preprocessing step or simultaneously with the optimization problem \citep{berg_data-driven_2019,both_deepmod_2019,chen_deep_2020}. Here, we preprocess the noisy displacements by denoising using kernel ridge regression (KRR). The noisy displacements in $\calU$ are spatially interpolated by a  ridge estimator kernelized by an RBF. The interpolation reduces the noise and the interpolated displacements are used as inputs to our algorithm. In \ref{sec:krr}, we briefly review KRR in the context of our work; for more details, the reader is referred to \cite{saunders_krr}. We specifically chose KRR as the denoiser because other methods are either computationally expensive (e.g., Gaussian process regression), only limited to rectangular domains with  structured discretizations (e.g., Gaussian filters), prone to overfitting (e.g., neural networks), or prone to underfitting (e.g., polynomial interpolation).

\renewcommand{\arraystretch}{1.2}
\begin{table}[!htb]
	\centering
	\caption{Strain energy density of the (true) hidden and discovered material models for different noise levels $\sigma$. GT$^\star$ denotes the case where the logarithmic feature is excluded from the feature library $\bfQ$.}%
	\label{tab:material_models}%
	\resizebox{\textwidth}{!}{%
		\begin{tabular}{lll}%
		\hline
			\multicolumn{2}{c}{Benchmarks} & \multicolumn{1}{l}{Strain Energy Density ($W$)} \\%
            \hline\\[-10pt]
			NH2 & Truth & $0.5000(\bar{I}_1-3) + 1.5000(J-1)^{2}$\\%
			~ & \CA $\sigma=0$ & \CA $0.5000(\bar{I}_1-3) + 1.5000(J-1)^{2}$\\%
			~ & \CB $\sigma=10^{-4}$ & \CB $0.4995(\bar{I}_1-3) + 1.4998(J-1)^{2}$\\%
			~ & \CC $\sigma=10^{-3}$ & \CC $0.4936(\bar{I}_1-3) + 1.4986(J-1)^{2}$\\%
			\\[-7pt] \hline\\[-10pt]
			NH4 & Truth & $0.5000(\bar{I}_1-3) + 1.5000(J-1)^{4}$\\%
			~ &\CA $\sigma=0$ & \CA $0.5000(\bar{I}_1-3) + 1.5000(J-1)^{4}$\\%
			~ &\CB $\sigma=10^{-4}$ &\CB $0.5005(\bar{I}_1-3) + 1.4973(J-1)^{4}$\\%
			~ &\CC $\sigma=10^{-3}$ &\CC $0.4848(\bar{I}_1-3) + 1.4728(J-1)^{4}$\\%
			\\[-7pt] \hline\\[-10pt]
			IH & Truth & $0.5000(\bar{I}_1-3) + 1.0000(\bar{I}_2-3) + 1.0000(\bar{I}_1-3)^2 + 1.5000(J-1)^{2}$\\%
			~ & \CA $\sigma=0$ &\CA $0.5000(\bar{I}_1-3) + 1.0000(\bar{I}_2-3) + 1.0000(\bar{I}_1-3)^2 + 1.5000(J-1)^{2}$\\%
			~ & \CB $\sigma=10^{-4}$ &\CB $0.5306(\bar{I}_1-3) + 0.9576(\bar{I}_2-3) + 0.9917(\bar{I}_1-3)^2 + 1.5041(J-1)^{2}$\\%
			~ & \CC $\sigma=10^{-3}$ &\CC $1.6323(\bar{I}_2-3) + 1.5546(\bar{I}_1-3)(\bar{I}_2-3) + 0.0304 (\bar{I}_1-3)^2(\bar{I}_2-3)^3 + 1.4498(J-1)^{2}$\\%
			\\[-7pt] \hline\\[-10pt]
			HW & Truth & $0.5000(\bar{I}_1-3) + 1.0000(\bar{I}_2-3) + 0.7000(\bar{I}_1-3)(\bar{I}_2-3) + 0.2000(\bar{I}_1-3)^3 + 1.5000(J-1)^{2}$\\%
			~ & \CA $\sigma=0$ &\CA $0.5000(\bar{I}_1-3) + 1.0000(\bar{I}_2-3) + 0.7000(\bar{I}_1-3)(\bar{I}_2-3) + 0.2000(\bar{I}_1-3)^3 + 1.5000(J-1)^{2}$\\%
			~ &\CB $\sigma=10^{-4}$ &\CB $0.5853(\bar{I}_1-3) + 0.9101(\bar{I}_2-3) + 0.6475(\bar{I}_1-3)(\bar{I}_2-3) + 0.2089(\bar{I}_1-3)^3 + 1.5050(J-1)^{2}$\\%
			~ &\CC $\sigma=10^{-3}$ & \CC $1.3600(\bar{I}_1-3) + 1.4284 (\bar{I}_2-3)^3 + 1.5469(J-1)^{2}$\\%
			\\[-7pt] \hline\\[-10pt]
			GT & Truth & $0.5000(\bar{I}_1-3) + 1.5000(J-1)^{2} + 1.0000 \log(\bar{I}_2/3)$\\%
			~ & \CA$\sigma=0$ &\CA $0.5000(\bar{I}_1-3) + 1.5000(J-1)^{2} + 0.9999 \log(\bar{I}_2/3)$\\%
			~ & \CB$\sigma=10^{-4}$ &\CB $0.4978(\bar{I}_1-3) + 1.5002(J-1)^{2} + 1.0095 \log(\bar{I}_2/3)$\\%
			~ & \CC$\sigma=10^{-3}$ &\CC $0.3578(\bar{I}_1-3) + 0.4695(\bar{I}_2-3) + 1.5152(J-1)^{2}$\\%
			\\[-7pt] \hline\\[-10pt]
			GT$^\star$ & Truth & $0.5000(\bar{I}_1-3) + 1.5000(J-1)^{2} + 1.0000 \log(\bar{I}_2/3)$\\%
			~ & \CA$\sigma=0$ &\CA $0.4105(\bar{I}_1-3) + 0.3783(\bar{I}_2-3) + 1.5040(J-1)^{2}$\\%
			~ & \CB$\sigma=10^{-4}$ &\CB $0.3974(\bar{I}_1-3) + 0.4011(\bar{I}_2-3) + 1.5045(J-1)^{2}$\\%
			~ & \CC$\sigma=10^{-3}$ &\CC $0.3578(\bar{I}_1-3) + 0.4695(\bar{I}_2-3) + 1.5152(J-1)^{2}$\\%
		\\[-7pt] \hline
		\end{tabular}%
	}%
\end{table}%
\renewcommand{\arraystretch}{1.0}

\begin{figure}[!ht]
	\begin{center}
		\includegraphics[height=0.7\textheight]{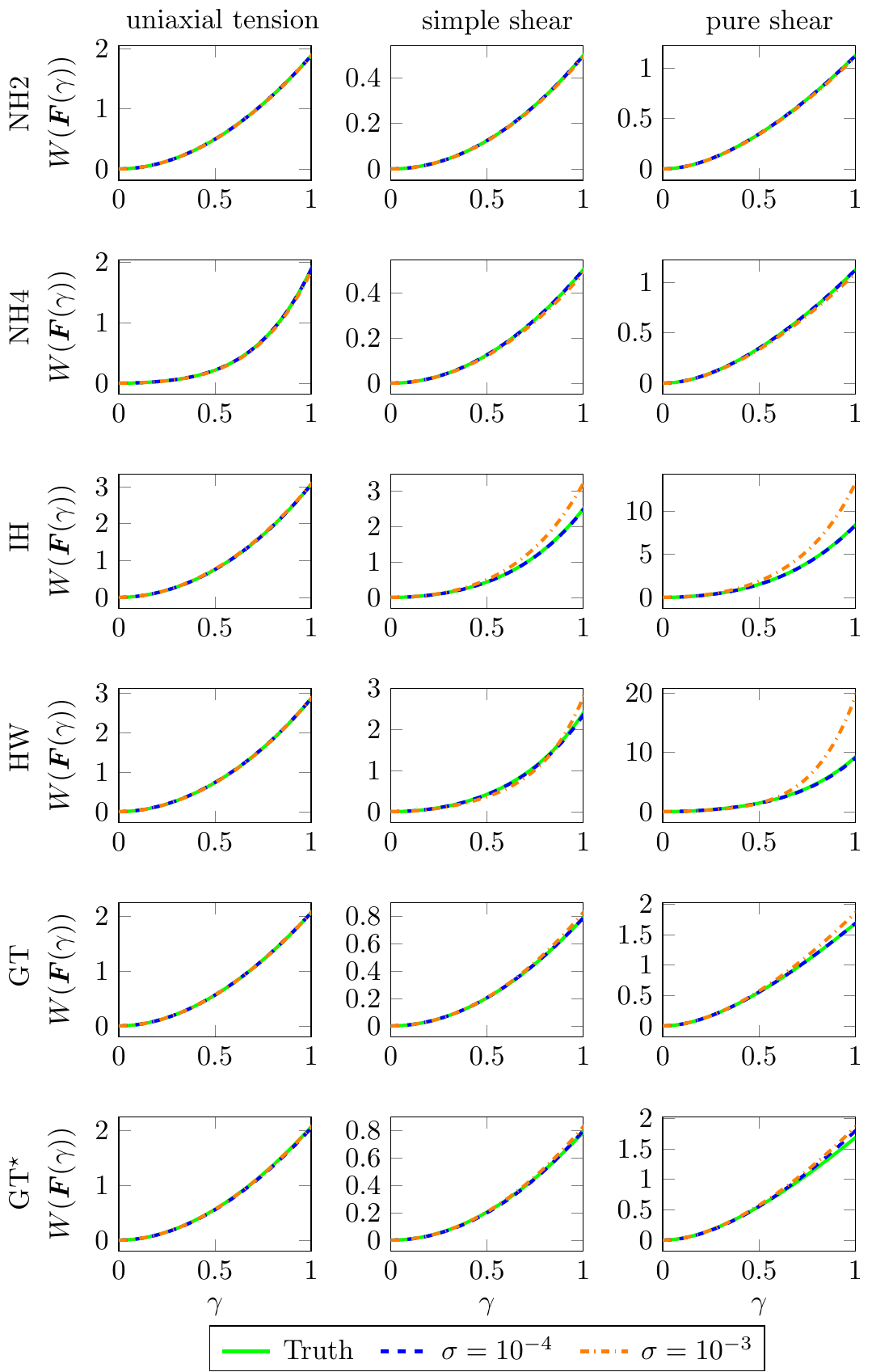}
		\caption{Strain energy density $W(\bm{F}(\gamma))$ along different deformation paths (see \eqref{eq:deformation_paths}) of the (true) hidden and discovered material models for different noise levels $\sigma$. The discovered models for the $\sigma=0$ case are omitted as they are exactly equal to the hidden models. GT$^\star$ denotes the case when the logarithmic feature is excluded from the feature library $\bfQ$.}
		\label{fig:strain_energy_density_plot}
	\end{center}
\end{figure}

\begin{figure}[!ht]
	\begin{center}
		\includegraphics[height=0.7\textheight]{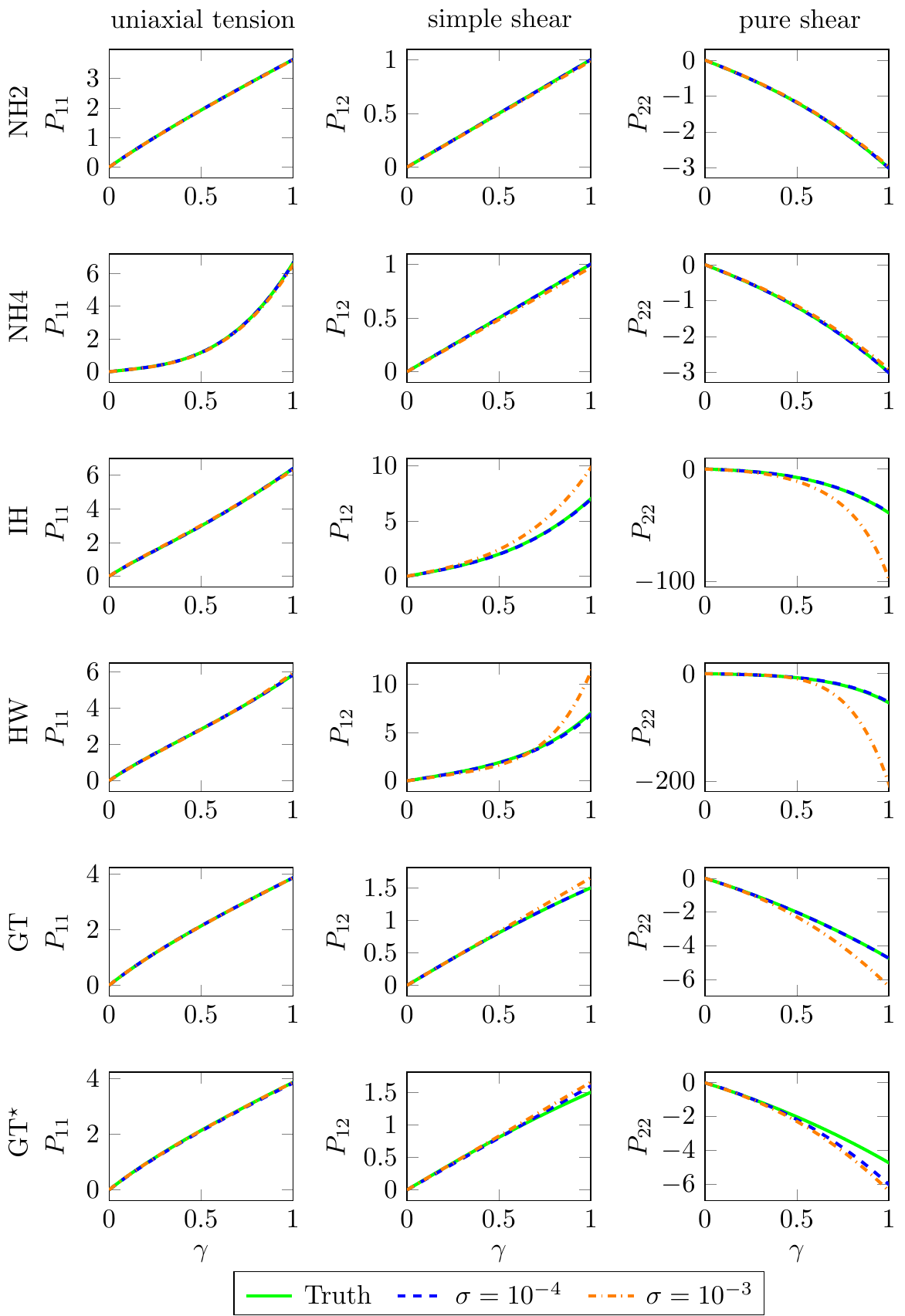}
		\caption{Stress components $P_{11}(\bm{F}^{\text{UT}}(\gamma))$ along uniaxial tension deformation path, $P_{12}(\bm{F}^{\text{SS}}(\gamma))$ along simple shear deformation path and $P_{22}(\bm{F}^{\text{PS}}(\gamma))$ along pure shear deformation path (see \eqref{eq:deformation_paths}) of the (true) hidden and discovered material models for different noise levels $\sigma$. The discovered models for the $\sigma=0$ case are omitted as they are exactly equal to the hidden models. GT$^\star$ denotes the case when the logarithmic feature is excluded from the feature library $\bfQ$.}
		\label{fig:stress_plot}
	\end{center}
\end{figure}

\subsection{Results}
\label{sec:results}

All subsequent results are based on the parameters (including the algorithm hyperparameters) presented in \ref{sec:settings}, which were identified upon a careful testing phase. The strain energy densities discovered by the proposed unsupervised algorithm from the data generated with the five chosen material models are reported in \tablename\ref{tab:material_models}. A comparison of the strain energy densities of the true models and the  discovered  models along three of the deformation paths in~\eqref{eq:deformation_paths} -- in particular, uniaxial tension (UT), simple shear (SS) and pure shear (PS) -- is presented in  \figurename\ref{fig:strain_energy_density_plot}. \figurename\ref{fig:stress_plot} further shows such a comparison for selected stress components, which can be deduced from the strain energy density by differentiating~\eqref{eq:stress}. As follows, these results are discussed for the three different noise levels ($\sigma$) artificially superposed to the displacement measurements. 

In the case of displacements without noise $(\sigma=0)$, all the material models are exactly discovered with negligible errors in the magnitude of the material parameters. In the low noise case ($\sigma = 10^{-4}$), the material models are again correctly discovered  (i.e.,  all the correct features are identified). There are small differences in the material parameters, which however  do not affect the  accuracy of the strain energy density. This conclusion is corroborated by the negligible difference in the plots of the strain energy densities and stress components along the chosen deformation paths in Figures~\ref{fig:strain_energy_density_plot} and~\ref{fig:stress_plot}.

In the high noise case ($\sigma = 10^{-3}$), the discovered models show good agreement with the hidden models for the two Neo-Hookean solids (NH2 and NH4). However, the unsupervised algorithm fails to correctly identify the model features for the Isihara (IH), Haines-Wilson (HW) and Gent-Thomas (GT) solids. Despite the difference in the functional form, the discovered  model for the GT solid shows good agreement with the true model in Figures~\ref{fig:strain_energy_density_plot} and~\ref{fig:stress_plot}. However, the discovered models for the IH and HW solids deviate from the true models -- particularly at large pure shear deformations. Hence, in these cases, they only serve as good approximations in the small deformation regime. This  also reveals the limitations in the extrapolatory capability of our approach in the case of high noise levels. Specifically, data with higher noise levels require larger displacements and loadings to improve the generalizability of the discovered models to larger deformations. This is also expected because the higher-order and nonlinear features are more dominant at larger deformations only. Moreover, the chosen experiment does not feature very large levels of pure shear deformations, which explains why the discovered model for higher noise deviates from the true model especially along this deformation path. Using experiments with more complex strain fields\footnote{See \cite{grediac_virtual_2008} and references therein for a variety of possible experimental setups.}, or using more than one experiment, is expected to be beneficial in cases where high noise levels are to be expected. 

The question arises whether the unsupervised  algorithm generalizes well when the feature library cannot represent the hidden material model due to the lack of the corresponding feature(s). To this end, we purposefully exclude the logarithmic feature $\log(\bar{I}_2/3)$ from the feature library $\bfQ$ ($n_f = 42$ in this case). Interestingly, in this situation the discovered  models for the GT solid, denoted by GT$^\star$ (\tablename\ref{tab:material_models} and Figures~\ref{fig:strain_energy_density_plot} and~\ref{fig:stress_plot}), surrogate the logarithmic feature by its linearization, i.e., by the feature $(\bar{I}_2-3)$. This is also observed for the high noise level ($\sigma = 10^{-3}$), for which it also occurs when the logarithmic feature is included in the feature library. The strain energy density and stress response is nevertheless reproduced quite accurately (Figures~\ref{fig:strain_energy_density_plot} and~\ref{fig:stress_plot}). These results suggest that the proposed approach is able not only to identify but also to surrogate model features, depending on the availability of such features in the catalogue of candidate functions, in such a way that the predicted response of the model is accurately reproduced.

\section{Conclusions and outlook}

We developed a new unsupervised sparse regression approach
for automated discovery of constitutive models for hyperelastic
material behavior. The approach requires only measurable (displacement and global force) data, 
fulfils many important physics constraints a priori, and delivers interpretable models from
a potentially very large library of candidate functions by enforcing the satisfaction of linear momentum balance in the interior and at the boundary of the domain. In order to discover parsimonious models with as little non-zero terms as possible, we adopt a sparsity promoting technique based on the classical $\ell_p$ regularization and, unlike in the previous literature, we determine the corresponding penalty parameter automatically through the enforcement of additional physics constraints. The devised algorithm combines this regularization with thresholding in a fully automatic fashion.
In our numerical experiments, where the datasets were obtained by finite element simulations with or without added artificial noise, the approach was shown to be able to correctly identify  five chosen hyperelastic models out of a large catalogue of functions based on only one loading experiment (i.e., in machine learning jargon, in one shot), with an accuracy depending on the magnitude of the noise and on the complexity of the model. For no or low noise, the models were identified perfectly or to high accuracy. For high noise, very accurate results were still obtained for the simplest models whereas the most complex ones were surrogated by others with reduced accuracy. In such cases, it is envisioned that loading experiments with more complex strain states and larger deformation levels as well as the availability of datasets from different experiments will be of help. Finally, we also showed that, if a ``true'' feature is missing in the library of candidate functions, the proposed approach is able to surrogate it in such a way that the actual response is still accurately predicted. 

We believe that the proposed approach has the potential for many future developments, including the employment on experimental data\footnote{Depending on the experimental setting, the algorithmic framework may need to be reformulated considering the plane stress assumption.} in the two- and three-dimensional settings and the extension to the more complex cases of non-homogeneous, anisotropic, and irreversible material behavior.

\appendix

\section{Denoising via kernel ridge regression (KRR)}
\label{sec:krr}

Here, we provide a brief review of KRR in the context of this work. For more details, the reader is referred to \cite{saunders_krr}.

Let $\bfy = \left(y^1,y^2,\dots,y^{n_n}\right)^T \in \Rset^{n_n}$ denote a vector of noisy scalar measurements, where $y^a$ is the measurement at the node of reference coordinate $\bfX^a\in\calX$. We assume an ansatz for the denoised field $\hat y:\Omega\rightarrow \Rset$ as
\be\label{eq:krr_ansatz}
\hat y(\bfX) = \bfvarphi(\bfX)^T\bfbeta,
\ee
where $\bfvarphi:\Omega\rightarrow\Rset^d$ denotes a highly nonlinear map to the $d$-dimensional feature space and $\bfbeta\in\Rset^d$ performs a linear combination of the rows of $\bfvarphi$. Here, $d$ is assumed to be much higher than the number of measurements, i.e., $d\gg n_n$, and can even be infinite. 

The objective is to determine the coefficients $\bfbeta$ such that the differences between noisy and denoised measurements are minimized subject to  $\ell_2$ regularization (also known as ridge regularization), i.e.,
\be\label{eq:krr_obj}
\bfbeta^\text{opt} =  \arg \min_{\bfbeta}  \| \bfPhi \bfbeta - \bfy \|^2 + \xi \|\bfbeta\|^2,\qquad  \text{with} \qquad \bfPhi = \left[
\bfvarphi(\bfX^1),\bfvarphi(\bfX^2),\dots,\bfvarphi(\bfX^{n_n})
\right]^T \in\Rset^{n_n\times d}.
\ee
Here, $\xi>0$ controls the strength of the $\ell_2$ regularization. The first optimality condition of \ref{eq:krr_obj} yields
\be\label{eq:krr_ridge}
\bfbeta^\text{opt} = \underbrace{\left( \xi \bfI + \bfPhi^T\bfPhi \right)^{-1} }_{d\times d}  \bfPhi^T \bfy.
\ee
This involves inverting a $d\times d$ matrix, which is computationally intractable if $d$ is infinite. 

To make \eqref{eq:krr_ridge} tractable, we perform the following manipulation,
\be
\bfbeta^\text{opt} =\left( \xi \bfI + \bfPhi^T\bfPhi \right)^{-1}  \bfPhi^T \underbrace{\left( \xi \bfI + \bfPhi\bfPhi^T \right)\left( \xi \bfI + \bfPhi\bfPhi^T \right)^{-1}}_{=\bfI}  \bfy = \left( \xi \bfI + \bfPhi^T\bfPhi \right)^{-1}   \left( \xi \bfPhi^T + \bfPhi^T\bfPhi\bfPhi^T \right)\left( \xi \bfI + \bfPhi\bfPhi^T \right)^{-1}  \bfy.
\ee
This is rearranged into
\be
\bfbeta^\text{opt}=\left( \xi \bfI + \bfPhi^T\bfPhi \right)^{-1}   \left( \xi \bfI + \bfPhi^T\bfPhi \right)\bfPhi^T\left( \xi \bfI + \bfPhi\bfPhi^T \right)^{-1}  \bfy,
\ee
which further simplifies to
\be\label{eq:krr_reduced}
\bfbeta^\text{opt}=\bfPhi^T \underbrace{\left( \xi \bfI + \bfPhi\bfPhi^T \right)^{-1}}_{n_n\times n_n}  \bfy.
\ee
Notably, \eqref{eq:krr_ridge} reduces to inverting a finite-dimensional $n_n\times n_n$ matrix which is computationally tractable in contrast to a $d\times d$ matrix (recall, $d\gg n_n$).

Substituting \eqref{eq:krr_reduced} into the ansatz \eqref{eq:krr_ansatz},  the denoised field is obtained as
\be
\hat y (\bfX) = \bfvarphi(\bfX)^T\bfbeta^\text{opt} = \bfvarphi(\bfX)^T \bfPhi^T \left( \xi \bfI + \bfPhi\bfPhi^T \right)^{-1} \bfy.
\ee
We then use the \textit{kernel trick} \citep{Theodoridis-PR}. For $\bfX,\tilde\bfX\in\Omega$, let
\be
\calK(\bfX,\tilde\bfX) = \bfvarphi(\bfX)^T\bfvarphi(\tilde\bfX),
\ee
be a positive definite kernel defining the inner product in the feature space $\varphi$. The denoised field is then reduced to
\be\label{eq:krr_kernelized}
\begin{aligned}
\hat y(\bfX) = &\bfk(\bfX)^T \left(\xi \bfI + \bfK \right)^{-1}\bfy,\\
\text{with}\quad  & k_a(\bfX) = \calK(\bfX,\bfX^a)
\quad\text{and} \quad K_{ab} = \calK(\bfX^a,\bfX^b),\quad \forall\quad a,b\in \{1,\dots,n_n\}.
\end{aligned}
\ee
The trick is that we only need to define a valid kernel $\calK$ (must satisfy the requirements of an inner product) to solve for the denoised field, without actually ever defining the (possibly infinitely-dimensional) feature space $\varphi$. To this end, we choose the RBF kernel defined as
\be
\calK(\bfX,\tilde\bfX) = \exp\left(\frac{-\|\bfX-\tilde\bfX\|^2}{2\chi^2}\right),
\ee
where $\chi>0$ controls the lengthscale of the kernel. Since KRR does not assume a parametric ansatz ($\bfvarphi$ remains undefined in \eqref{eq:krr_ansatz}), it is a \textit{non-parametric} regression.

In the context of this work, the displacement components are denoised separately by setting $\bfy=(u^1_1,u^2_1,\dots,u^{n_n}_1)^T$ and $\bfy=(u^1_2,u^2_2,\dots,u^{n_n}_2)^T$ in \eqref{eq:krr_kernelized}. The only hyperparameters are the  regularization strength ($\xi$) and kernel length scale ($\chi$). These hyperparameters are optimized by minimizing the error $\sum_{a=1}^{n_n}\|\hat y(\bfX^a)-y^a\|^2$  via random search.

We note that KRR is equivalent to estimating the posterior mean in Gaussian process regression \citep{Rasmussen-GPR,gundersen_2020}. However, the latter also has additional complexity as it provides uncertainties and  posterior sampling for the predictions $\hat\bfy$ (unlike the deterministic framework of KRR). Consequently, despite the asymptotic computational complexity of both regressions being $O(n_n^3)$, we observed that KRR is significantly faster for moderately-sized problem ($n_n\sim 10^4-10^5$) in the context of this work.

\begin{table}[htb]
	\caption{Default parameters and hyperparameters for the data generation and the proposed algorithm.}\label{tab:settings}
	\centering
	\begin{tabular}{lcc}
		\hline
		Parameter & Notation & Value \\ \hline
		Number of nodes in mesh & $n_n$ & $63,601$ \\
		Number of reaction force constraints & $n_\alpha$ & $4$ \\
		Number of loadsteps & $L$ & NH2, NH4: 4 \\ & & \ \ IH, HW, GT: 8 \\
		Loading parameter & $\delta$ & $\left\{0.1\times l : l=1,\dots ,L\right\}$ \\
		Number of features & $n_f$ & $\left\{42,43\right\}$ \\
		Coefficient for reaction force balance & $\lambda_r$ & 100 \\
		$\ell_p$ regularization & $p$ & $1/4$ \\
		Initial coefficient for $\ell_p$ regularization & $\lambda^0_p$ & $0.01$ \\
		Multiplicative factor for increasing $\lambda_p$ & $\kappa$ & 5 \\
		Number of parallel fixed-point iterations & $-$ & $200$ \\
		Maximum number of fixed-point iteration steps & $-$ & $200$ \\
		Number of samples for admissibility checks & $n_\gamma$ & $75$\\
		Maximum loading parameter for admissibility checks & $\gamma_\text{max}$ & $10^{9}$ \\
		Zero tolerance for $\bftheta$ during fixed-point iteration & $\epsilon_\text{tol}$ & $10^{-6}$ \\
		Convergence tolerance for $\bftheta$ during fixed-point iteration & $\epsilon_\text{conv}$ & $10^{-3}$\\
		Threshold for $\bftheta$ after fixed-point iteration & $\bar\theta$ & $0.01$ \\
		Displacement noise standard deviation & $\sigma$ & $\left\{0,10^{-4},10^{-3}\right\}$\\ \hline
	\end{tabular}
\end{table}

\section{Numerical protocol and computational costs}
\label{sec:settings}

Table~\ref{tab:settings} lists the default set of parameters and hyperparameters used in the data generation and the proposed algorithm. For the NH2 and NH4 models, displacement data from  $L=4$ loadsteps are considered. For the rest of the models, i.e., IH, HW, and GT, with higher order terms in the strain energy density, the number of considered load steps is increased to $L=8$ to allow for similar maximum strain levels as in the simpler models.

For the chosen number of nodes ($n_n$) in the spatial discretization, the computational time needed to execute the algorithm depicted in \figurename\ref{fig:sparsity_promotion_flowchart} with one initial guess for the fixed-point iterations is of the order of ten minutes for an average modern processor. We highlight that the processes of fixed-point iterations with different initial guesses can be executed in an embarrassingly parallel fashion on a distributed cluster.

\section*{Code and data availability}

The codes and data generated during the current study are available from the authors upon reasonable request.

\bibliographystyle{elsarticle-harv}
\bibliography{Bib}

\end{document}